\documentclass[12pt,a4paper]{article}
\usepackage{amsmath,amssymb}
\usepackage{url,cite,slashed}
\usepackage{graphicx}
\usepackage{subfigure}

\makeatletter
    
    \@addtoreset{equation}{section}
\makeatother

\newcommand{\sss}{\tilde{s}}
\newcommand{\nn}{\nonumber}
\newcommand{\simlt}{\lower.5ex\hbox{$ \buildrel < \over \sim $}}
\newcommand{\simgt}{\lower.8ex\hbox{$ \buildrel > \over \sim $}}

\newcommand{\invfb}{{\rm fb}^{-1}}
\newcommand{\brinv}{{\rm BR}_{\rm inv}}
\newcommand{\clim}{c_\chi^{\rm lim}}

\begin{document}

\renewcommand{\thefootnote}{\fnsymbol{footnote}} % For titlepage
\begin{titlepage}

\begin{center}

\hfill UT--14--33\\

\vskip .75in

{\Large \bf 
Heavy WIMP through Higgs portal at the LHC
}

\vskip .75in

{\large
Motoi Endo
and 
Yoshitaro Takaesu
}

\vskip 0.25in

{\em Department of Physics, University of Tokyo, Tokyo 113--0033, Japan}

\end{center}

\vskip .5in

\begin{abstract}

LHC constraints on Higgs-portal WIMPs are studied.
Scalar, vector and anti-symmetric tensor fields are considered. 
They are assumed to be heavier than a half of the Higgs boson mass. 
We investigate 8\,TeV LHC results on signatures of the vector boson fusion, 
mono-jet and associated production of the $Z$ boson, which proceed via
virtual exchange of the Higgs boson. 
We show that Higgs-portal interactions of the vector and tensor WIMPs are 
constrained to be less than 0.43 and 0.16, respectively, while those for 
the scalar are very weak. 
Prospects of the 14\,TeV LHC are also discussed.

\end{abstract}

\end{titlepage}

\setcounter{page}{1}
\renewcommand{\thefootnote}{\#\arabic{footnote}}
\setcounter{footnote}{0}

%%%%%%%%%%%%%%%%%%%%%%%%%%%%%%%%%%%%%%%%%%%%%%%%%%%%
\section{Introduction}
Although astrophysical and cosmological evidences have established the existence of the dark matter (DM), its nature has not been demystified (see, e.g., Ref.~\cite{Bertone:2004pz} for a review). 
In particular, DM interactions with standard model (SM) particles have not been identified except for gravitational ones in spite of enormous experimental efforts.
They have been limited by direct or indirect DM searches, for instance, by the LUX experiment \cite{Akerib:2013tjd}. 
Besides, if the DM has sizable couplings with quarks or gluons, it
can be discovered or constrained by collider experiments especially at
the LHC \cite{Eboli:2000ze}. 
In fact, searching for the DM is one of the main targets for the next phase
of the LHC.

The collider searches are qualitatively different from the direct and
indirect ones for the DM. 
The latter particularly depends on the relic abundance of the DM as well as its interactions. 
For a subcomponent of the DM, the signal strength of direct or indirect searches can be suppressed, even if its coupling is strong.
In contrast, signal strengths at colliders depend only on interactions. 
Weakly-interacting massive particles (WIMPs) are searched, irrespective of whether they are dominant components of the DM.
They are assumed to be stable and identified as a missing momentum in detectors.

In this letter, we study LHC signatures of stable WIMPs through the Higgs portal, where they interact with the SM particles only via the Higgs boson.
They are assumed to be singlet under the SM gauge symmetries. 
An unbroken $Z_2$ parity is introduced, where the stability of WIMPs is guaranteed by assigning odd (even) charge to WIMP (SM) particles. 
We consider scalar, vector and anti-symmetric tensor fields as a candidate of Higgs-portal WIMPs\footnote
{
The stability of the tensor WIMP does not always require a new $Z_2$ symmetry \cite{Cata:2014sta}.
}.

Although the Higgs-portal models have been constrained by LHC studies on the Higgs invisible decay, they target WIMPs lighter than a half of the Higgs boson mass\cite{Aad:2014iia,Chatrchyan:2014tja,Belanger:2013kya}. 
In this letter, we study LHC signatures of WIMPs when they are {\it heavier}.
We explore the vector boson fusion (VBF) and $Z$-boson associated production channels. 
Besides, WIMPs are produced by gluon fusions via top loops and the intermediate Higgs boson. 
Such a channel can be identified by using associate productions of a hard jet.
Thus, we also investigate the mono-jet signature in light of the current LHC studies\cite{ATLAS:2012zim,CMS:rwa}.
It will be shown that the current LHC results constrain the Higgs-portal
interactions of the vector and tensor WIMPs to be less than 0.43 and 0.16, respectively, while those for the scalar is very weak. 
Prospects of the 14\,TeV LHC will also be discussed.

The remaining parts of this work are organized as follows. After
introducing the Higgs-portal models in Section \ref{sec:Model}, the
relevant LHC signatures and analysis details are described in Section \ref{sec:signature}. Results
for the 8\,TeV
LHC constraints are shown in Section \ref{sec:limits}, while Section \ref{sec:prospects} is devoted to discussions on prospects for the 14\,TeV LHC. Finally, our main conclusion are summarized in Section \ref{sec:conclusion}. 

%%%%%%%%%%%%%%%%%%%%%%%%%%%%%%%%%%%%%%%%%%%%%%%%%%%%
\section{Model}
\label{sec:Model}
Higgs-portal WIMP models are considered, where WIMP is a scalar,
$S$, vector, $V_\mu$, or anti-symmetric tensor field, $B_{\mu\nu}$\footnote
{
It is straightforward to apply the following analysis for a fermion WIMP, which interacts with the Higgs boson by a dimension-five operator. 
}.
It generally couples to the Higgs boson via a dimension-four interaction operator. 
The Lagrangian is generally given as
\begin{subequations}
\begin{align}
  \mathcal{L}_S &= 
  \frac{1}{2} \partial^\mu S \partial_\mu S - \frac{1}{2} M_S^2 S^2
   - \lambda_S S^4 - c_S \left| H \right|^2 S^2, \\
  \mathcal{L}_V &= 
  - \frac{1}{4} V^{\mu\nu}V_{\mu\nu} + \frac{1}{2} M_V^2 V^\mu V_\mu
   - \lambda_V (V^\mu V_\mu)^2 + c_V \left| H \right|^2 V^\mu V_\mu, \\
  \mathcal{L}_B &= 
  \frac{1}{4} \partial_\lambda B^{\mu\nu} \partial^\lambda B_{\mu\nu}
  - \frac{1}{2} \partial^\mu B_{\mu\nu} \partial_\rho B^{\rho\nu}
  - \frac{1}{4} M_B^2 B^{\mu\nu} B_{\mu\nu} \notag \\
  & ~~~
  - \lambda_B B_{\mu\nu} B^{\nu\lambda} B_{\lambda\rho} B^{\rho\mu}
  - c_B \left| H \right|^2 B^{\mu\nu} B_{\mu\nu},
\end{align}
\label{eq:model}
\end{subequations}
where $M_\chi$, $\lambda_\chi$ and $c_\chi$ are the mass parameter, quartic self-coupling and interaction strength between the Higgs boson and WIMP, respectively, for $\chi = S, V, B$.
In the LHC analysis, $c_\chi$ is set to be a real and positive value without loss of generality. 
Also, $V_{\mu\nu}$ is a field strength of the vector.
After the electroweak symmetry is broken, the WIMP mass receives a correction of $\sim c_\chi v^2$.
In the analysis, the physical WIMP mass is represented as $m_\chi$.
In this letter, we follow Ref.~\cite{Cata:2014sta} for the tensor
model, where a massive anti-symmetric two-form field is analyzed in the
transverse representation as a candidate of the DM.
In particular, the wave function becomes
\begin{align}
  \left\langle 0 \left| B_{\mu\nu} \right| b(q, \lambda) \right\rangle =
  \frac{i}{m_B} \epsilon_{\mu\nu\rho\sigma}\, \varepsilon^\rho(\lambda) q^\sigma,
  \label{eq:wavefunction}
\end{align}
for momentum $q$ and helicity $\lambda$.

%%% partial widths for h > SS, h > VV and h > TT
The invisible decay rate of the Higgs boson with a mass $m_H$ is calculated as
\begin{subequations}
\begin{align}
  \Gamma_S(m_H,m_S;c_S) &= \frac{c_S^2}{8\pi} \frac{v^2}{m_H} \sqrt{ 1 - \frac{4m_S^2}{m_H^2} }, \\
  \Gamma_V(m_H,m_V;c_V) &= \frac{c_V^2}{32\pi} 
  \frac{v^2}{m_H} \frac{m_H^4 - 4 m_H^2 m_V^2 + 12 m_V^4}{m_V^4} 
  \sqrt{ 1 - \frac{4m_V^2}{m_H^2} }, \\
  \Gamma_B(m_H,m_B;c_B) &= \frac{c_B^2}{4\pi} 
  \frac{v^2}{m_H} \frac{m_H^4 - 4 m_H^2 m_B^2 + 6 m_B^4}{m_B^4} 
  \sqrt{ 1 - \frac{4m_B^2}{m_H^2} },
\end{align}
 \label{eq:width}
\end{subequations}
where $v$ is the vacuum expectation value of the Higgs field, $v \simeq 246\,{\rm GeV}$.
It is noticed that the vector and tensor productions are enhanced by $m_H^4/m_\chi^4$, if $m_H$ is much larger than $m_\chi$.
For the vector final state, this comes from the longitudinal polarization.
On the other hand, in the tensor case the wave function
\eqref{eq:wavefunction} is proportional to
$\varepsilon^{\rho}q^\sigma/m_B$ but vanishes for
$\varepsilon^\rho(\lambda) \propto q^\rho$, leading to the similar
scaling behavior in high energy as the vector case. 
Consequently, the Higgs invisible decay rate becomes large especially for the vector and tensor WIMPs.

%%%%%%%%%%%%%%%%%%%%%%%%%%%%%%%%%%%%%%%%%%%%%%%%%%%%
\section{LHC Signatures}
\label{sec:signature}
In this section we discuss LHC signatures for the Higgs-portal
 models described in the previous section. 

Higgs invisible decay signals are definitive probes for the
Higgs-portal interactions. ATLAS and CMS collaborations
have put limits on the branching fraction of the Higgs invisible decay 
($\brinv$) based on VBF and $ZH$ associated
production processes\cite{Aad:2014iia,Chatrchyan:2014tja}. 
Those constraints can be reinterpreted for bounds on the heavy Higgs-portal WIMP models, 
where off-shell Higgs bosons intermediate between the SM particles
and WIMPs. 
Besides, new physics searches via mono-jet
\cite{CMS:rwa,ATLAS:2012zim}, mono-$Z$\cite{Aad:2013oja,Aad:2014vka} and
mono-$W$\cite{Aad:2013oja,CMS:2013iea} signatures can also be a way to investigate Higgs-portal models.

In the following, we explain details of our VBF, mono-jet and
mono-$Z$
analyses for constraints on the heavy Higgs-portal models. On the other
hand, since the
currently available LHC results for the $ZH$\cite{Aad:2014iia,Chatrchyan:2014tja} and mono-$W$\cite{Aad:2013oja,CMS:2013iea} signatures 
are based on the template-based analyses (i.e., depending on kinematical
distributions of decay products), it is not straightforward to
reinterpret them for the constraints on the Higgs-portal
models. We do not investigate these channels in this study.

%%%%%%%%%%%%%%%%%%%%%%%%%%%%%%%%%%%%%%%%%%%%%%%%%%%%
\subsection{Vector Boson Fusion}
We briefly explain the analysis details for VBF constraints on the Higgs-portal models.
WIMP-pair productions are intermediated by the Higgs boson, $H^*$, as
\begin{equation}
pp \to H^* + jj \to \chi\chi + jj,
\end{equation}
where the Higgs is off-shell when the WIMP is heavier than
a half of the Higgs boson mass.
Its cross section can be expressed as
%% master formula for pp > j\chi\chi +X
\begin{equation}
 \sigma_{\chi\chi}(m_\chi,c_\chi) = \int_{4m_\chi^2}^\infty \frac{d\sss}{2\pi}\,
  \sigma_H(m_H = \sqrt{\tilde{s}})
  \frac{2\sqrt{\sss}}{(\sss -m_h^2)^2 +\Gamma_h^2 m_h^2}\,
  \Gamma_{\chi}(\sqrt{\sss},m_\chi; c_\chi),
\label{eq:xsec}
\end{equation}
where $\sss$ is the invariant mass squared of the $\chi\chi$ system, 
and $\Gamma_h = 4.21$\,MeV is the total decay width of the Higgs boson 
at the mass, $m_h = 125$\,GeV \cite{Heinemeyer:2013tqa}. 
The cross section consists of the following three parts: i) the production cross section 
of the Higgs boson, $\sigma_H$, at a hypothetical mass of the Higgs boson, $m_H = \sqrt{\tilde{s}}$,
ii) the Higgs boson propagator at the four momentum squared $\tilde{s}$, and iii) the decay 
width of the Higgs boson into a pair of WIMPs, $\Gamma_{\chi}$, provided in 
Eq.~\eqref{eq:width}, where the Higgs mass is supposed to be $m_H = \sqrt{\tilde{s}}$. 
It should be noted that the hypothetical Higgs boson mass $m_H$ is not equal to the SM Higgs boson mass $m_h$.
Higgs production cross sections, $\sigma_H$, depend on Higgs production processes, e.g., Higgs
plus one jet, VBF and
$ZH$ associated production, while the Higgs propagator and $\Gamma_{\chi}$ are universal for
all the production processes. Thus, our main concern in evaluating 
a WIMP-pair production cross section is the
calculation of the corresponding Higgs production cross section for various Higgs masses. 

The cross section, $\sigma_H$, for the VBF process is calculated with the HAWKv2.0 package\cite{Denner:2011id,HAWK} at the next-to-leading order (NLO), including QCD and EW corrections.
We impose the following kinematical cuts based on the CMS VBF analysis\cite{Chatrchyan:2014tja}: 
\begin{align}
 &p_{Tj_{1(2)}} > 50\,{\rm GeV},\,\,\, |\eta_{j_{1(2)}}| < 4.7, \nn\\
 &\eta_{j_1} \eta_{j_2} < 0,\,\,\, \Delta\eta_{jj} > 4.2,\,\,\, 
 |\Delta\phi_{jj}| < 1,\,\,\, M_{jj} > 1100\,{\rm GeV}, \\
 &p_{TH} > 130\,{\rm GeV},\nn
\end{align}
where $p_{T{j_{1(2)}}}$ and $\eta_{j_{1(2)}}$ are the transverse momentum and pseudo-rapidity
of the first (second) leading jet, respectively. Also, $\Delta\eta_{jj},
\Delta\phi_{jj}$ and $M_{jj}$ are the rapidity difference,
azimuthal-angle difference and invariant mass of the two leading jets,
respectively. Since the missing momentum arises solely from the Higgs 
invisible decay in our simulation, the Higgs transverse momentum, $p_{TH}$, 
is equal to the missing transverse momentum, $\slashed{p}_T$.
The CTEQ6L1 parton distribution function (PDF) set\cite{Pumplin:2002vw} is used with the factorization/renormalization scale equal 
to $(m_H^2+p_{TH{\rm cut}}^2)^{1/2}$, where $p_{TH{\rm cut}}$ is the Higgs $p_T$ cut.
The following analysis is based on the parton level calculation, and effects of 
parton shower, hadronization and detector simulation are neglected. 
Besides, we assume 100\% trigger and reconstruction efficiencies, supported by their sufficiently high performance \cite{Chatrchyan:2014tja}.
Also HAWK is modified to implement the $\Delta\phi_{jj}$ cut.

The CMS collaboration has reported that the observed 95\% confidence level (CL) 
upper bound on $\brinv$ is 0.65 for the SM Higgs boson with $m_h = 125$\,GeV, 
where the upper bound on the signal event number is 
$210 \times 0.65 \simeq 137$ for the integrated luminosity (${\cal L}$)
of $19.5\,\invfb$ at $\sqrt{s} = 8$\,TeV\cite{Chatrchyan:2014tja}.
If we require the VBF Higgs production cross section times the
invisible decay branching fraction, $\sigma_H(m_H)\times \brinv$, to be smaller than 
the CMS bound, $\sigma = 210 \times 0.65 / 19.5\,\invfb \simeq 7$\,fb,
we can reproduce the CMS upper bound on $\brinv$ as a function of the
Higgs mass (Fig.~8 in Ref.~\cite{Chatrchyan:2014tja}) within 
the $1\sigma$ level of the experimental uncertainty. 
Our bound is stronger than that of CMS. The difference may originate in
theoretical uncertainties from PDF and factorization/renormalization scales as well as the 
effects of parton shower, hadronization and detector smearing including the trigger and 
reconstruction efficiencies, which are neglected in our analysis. 

The WIMP production cross section, $\sigma_{\chi\chi}$, is calculated from Eqs.~\eqref{eq:xsec}. 
Contributions of lower $\sss$ dominates the integration.
The 95\% CL upper bound on the Higgs-portal coupling,
$c_{\chi}$, is obtained by requiring the expected event number from the
WIMP production process to be less than the CMS limit on the signal excess, 
137 events.

%%%%%%%%%%%%%%%%%%%%%%%%%%%%%%%%%%%%%%%%%%%%%%%%%%%%
\subsection{Mono jet}
Next we explain the analysis details for the mono-jet constraints on the Higgs-portal models. 
%%%% NLO_xsec = LO_xsec(finite mt) * K-factor(large mt)
The WIMP production process contributing to the mono-jet signal is 
\begin{equation}
pp \rightarrow
H^* + j \rightarrow \chi \chi + j,
\label{eq:monojet}
\end{equation}
where $j$ denotes an associated jet,
and the intermediate Higgs is off-shell when the WIMP is heavier than a half
of the Higgs boson mass. 

As in the VBF analysis in the previous subsection, the WIMP-pair production cross section is expressed by Eq.~\eqref{eq:xsec}.
Here, the Higgs production cross section, $\sigma_H$, corresponds to 
that of the Higgs plus one hard jet process at a hypothetical Higgs mass, $m_H = \sqrt{\tilde{s}}$. 
The production proceeds mainly via top quark loops.
When either the Higgs boson mass or the jet transverse momentum exceeds 
twice the top quark mass, $1/m_t$ expansion of the top loop
function breaks down, and the finite $m_t$ effects must be taken
into account. This is the case when a Higgs-portal WIMP is heavy.
The leading order (LO) cross section of the process has been evaluated with the finite top
mass\cite{Ellis:1987xu}. However, the NLO cross section
is known only in the infinite $m_t$ limit to the best of our
knowledge. Therefore, we take account of the finite $m_t$ effect at
the LO level and include the NLO corrections approximately by a K-factor,
defined as the ratio of the NLO and LO cross sections in the infinite
$m_t$ limit as
\begin{align}
 &\sigma_H^{\rm NLO} \simeq
   \sigma_{pp \to Hj}^{\rm LO}(m_t) \times K, \label{eq:monojet_xsec}\\
 &K = \frac{\sigma_{pp \to Hj}^{\rm NLO}(m_t=\infty)}{\sigma_{pp \to Hj}^{\rm LO}(m_t=\infty)},
\label{eq:K}
\end{align}
where the top quark mass is taken to be $m_t = 173\, {\rm GeV}$.
This procedure should involve potentially large uncertainties mainly due to the breakdown of the infinite $m_t$ approximation.
Therefore, it is desirable to calculate finite $m_t$ corrections to the NLO cross section.

The CMS collaboration has reported the most severe limit on the mono-jet 
production cross section by using the $19.5\,\invfb$ dataset at 
$\sqrt{s}=8$\,TeV \cite{CMS:rwa}, while the ATLAS collaboration puts a weaker limit based 
on the ${\cal L} = 10.5\,\invfb$ dataset \cite{ATLAS:2012zim}.
We follow the CMS analysis \cite{CMS:rwa} and impose the kinematical 
cut condition,
\begin{align}
 &p_{Tj_1} > 110\, {\rm GeV},\, |\eta_{j_1}| < 2.4, \notag \\
 &p_{TH} > 450\, {\rm GeV},
 \label{eq:cut-monojet}
\end{align}
where $p_{Tj_1}$ and $\eta_{j_1}$ are the transverse momentum and pseudo-rapidity
of the highest $p_{T}$ jet, respectively, and $p_{TH}$ is the transverse momentum
of the Higgs boson. 
The $p_{TH}$ cut is related to the $\slashed{p}_T$ cut since the Higgs boson decays
invisibly to WIMPs.
The above $p_{TH}$ cut condition is chosen to give the most stringent
limit on the Higgs-portal interactions, based on the CMS analysis \cite{CMS:rwa}. 
In our analysis, the cut conditions, Eq.~\eqref{eq:cut-monojet}, are imposed on $\sigma_{pp \to Hj}^{\rm LO}(m_t)$, whereas in the evaluation of the K-factor we take the $p_{TH}$ cut as low as $m_{H}/2$\footnote
{
This choice of the $p_{TH}$ cut avoids large $\log(m_H/p_{TH})$ corrections (see Ref.~\cite{Kauffman:1991cx}).
} instead of taking 450\,GeV in order to tame the large deviation from the finite $m_t$ result. 
Here, we suppose that the K-factor does not change significantly with $p_{TH}$ (cf.~Ref.\cite{deFlorian:1999zd}).
We use the modified MCFMv6.8 package\cite{MCFM}, where we implement the $p_{TH}$ cut,
to calculate the cross section. 
The LO (NLO) MSTW2008 PDF set\cite{Martin:2009iq} is used for the LO (NLO) cross section
calculations with the renormalization/factorization scale of $(m_{H}^2+p_{TH{\rm cut}}^2)^{1/2}$.
It is mentioned that the NLO corrections may be overestimated, since we do not impose the veto on the secondary jet from
the NLO real emission corrections unlike the CMS analysis\cite{Chatrchyan:2014tja}.

For $p_{TH} > 450\,{\rm GeV}$, which is equivalent to $\slashed{p}_T > 450\,{\rm GeV}$ 
at LO, the observed CMS upper bound on the signal event number is 157 at 95\% CL after imposing 
the cuts at $\sqrt{s}=8$\,TeV and ${\cal L}=19.5\,\invfb$ \cite{Chatrchyan:2014tja}. 
Thus, the production cross section of the mono-jet signal is required to be smaller than 
$\sigma = 157/19.5\,\invfb \simeq 8.1$\,fb at 95\% CL. 
This CMS result constrains the Higgs-portal coupling, $c_{\chi}$, in the similar way as the VBF case in the previous section.

%%%%%%%%%%%%%%%%%%%%%%%%%%%%%%%%%%%%%%%%%%%%%%%%%%%%
\subsection{Mono $Z$}
The WIMP production process contributing to the mono-$Z$ signal is 
\begin{equation}
pp \rightarrow
H^* + Z \rightarrow \chi \chi + Z.
\label{eq:monoZ}
\end{equation}
The ATLAS collaboration has reported dark matter searches with hadronically
decaying mono-$W/Z$ and leptonically decaying mono-$Z$ channels
with
$\mathcal{L} = 20.3\,\invfb$ at
$\sqrt{s}=8$\,TeV\cite{Aad:2013oja,Aad:2014vka}\footnote
{
Signal topologies of the $ZH$ production processes\cite{Aad:2014iia,Chatrchyan:2014tja}, where the Higgs boson is supposed to decay invisibly, are the same as those of the mono-$Z$ production processes\cite{Aad:2013oja,Aad:2014vka}. However, Refs.~\cite{Aad:2014iia,Chatrchyan:2014tja} uses the template-based analysis, and we do not consider them.
}. 
The most severe limit on the mono-$Z$
production cross section is set by the leptonic mode. We follow the
leptonic mode analysis\cite{Aad:2014vka} and impose the kinematical 
cut condition,
\begin{align}
 &p_{Tl} > 20\, {\rm GeV},\, |\eta_l| < 2.5, \notag \\
 &76\, {\rm GeV} < m_{ll} < 106\, {\rm GeV}, \notag \\
 &|\Delta\eta_{ll}| < 2.5 \notag, \\
 &p_{TH} > 150\, {\rm GeV},
 \label{eq:cut-monoZ}
\end{align}
where $p_{Tl}$ and $\eta_l$ are lepton transverse momentum and pseudo-rapidity
($l=e,\mu$), and $m_{ll}$ and $\Delta\eta_{ll}$
are the invariant mass and the difference of the pseudo-rapidities
of the two leading leptons, respectively. 
The above $p_{TH}$ cut condition is chosen to give the most stringent
limit on the Higgs-portal interactions, based on the ATLAS
analysis \cite{Aad:2014vka}. 
We use the HAWKv2.0 package\cite{Denner:2011id,HAWK}, where we have implemented
the $\Delta\eta_{ll}$ cut, and calculate the cross section of the Higgs production process, $pp
\rightarrow H^{*} Z$, including NLO QCD and EW contributions.  
The CTEQ6M PDF set\cite{Pumplin:2002vw} is used with 
the renormalization/factorization scale of $m_H+m_Z$. 
It is mentioned that we do not impose jet vetoes unlike the ATLAS
analysis\cite{Aad:2014vka}.

For $\slashed{p}_T > 150\,{\rm GeV}$, which is equivalent to $p_{TH} > 150\,{\rm GeV}$ 
at LO, the observed ATLAS upper bound on the production cross section is 2.7 $\invfb$ at 95\% CL.
This result constrains the Higgs-portal coupling, $c_{\chi}$, in
a similar way as the VBF and mono-jet cases.

%%%%%%%%%%%%%%%%%%%%%%%%%%%%%%%%%%%%%%%%%%%%%%%%%%%%
\section{Limits on the Higgs-portal models}
\label{sec:limits}
%%%%%%%%%%%%%%%%%%%%%%%%%%%%%%%%%%%%%%%%%%%%%%%%%%%%
\begin{figure}[tbp]
\begin{center}
\subfigure[]{
 \includegraphics[scale=0.6]{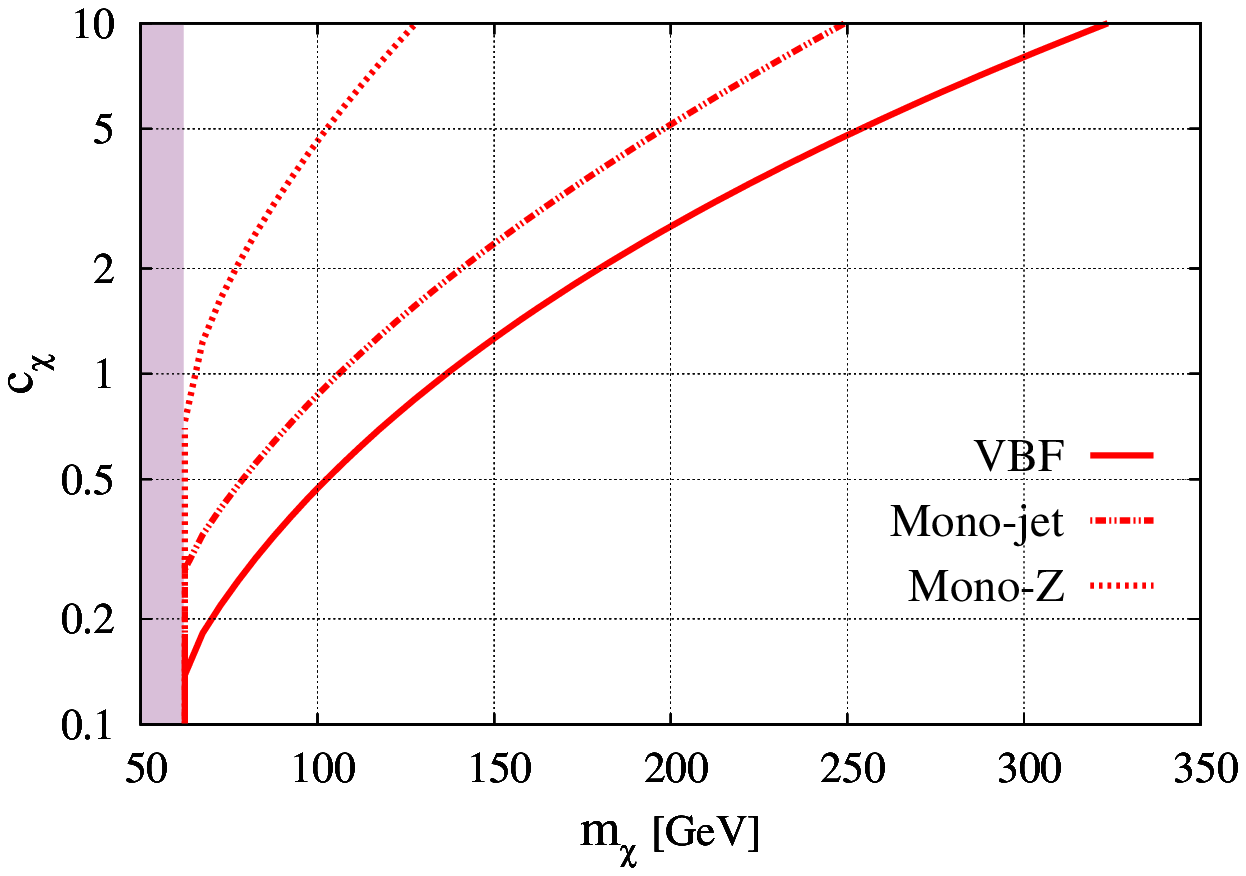}
}
\subfigure[]{
 \includegraphics[scale=0.6]{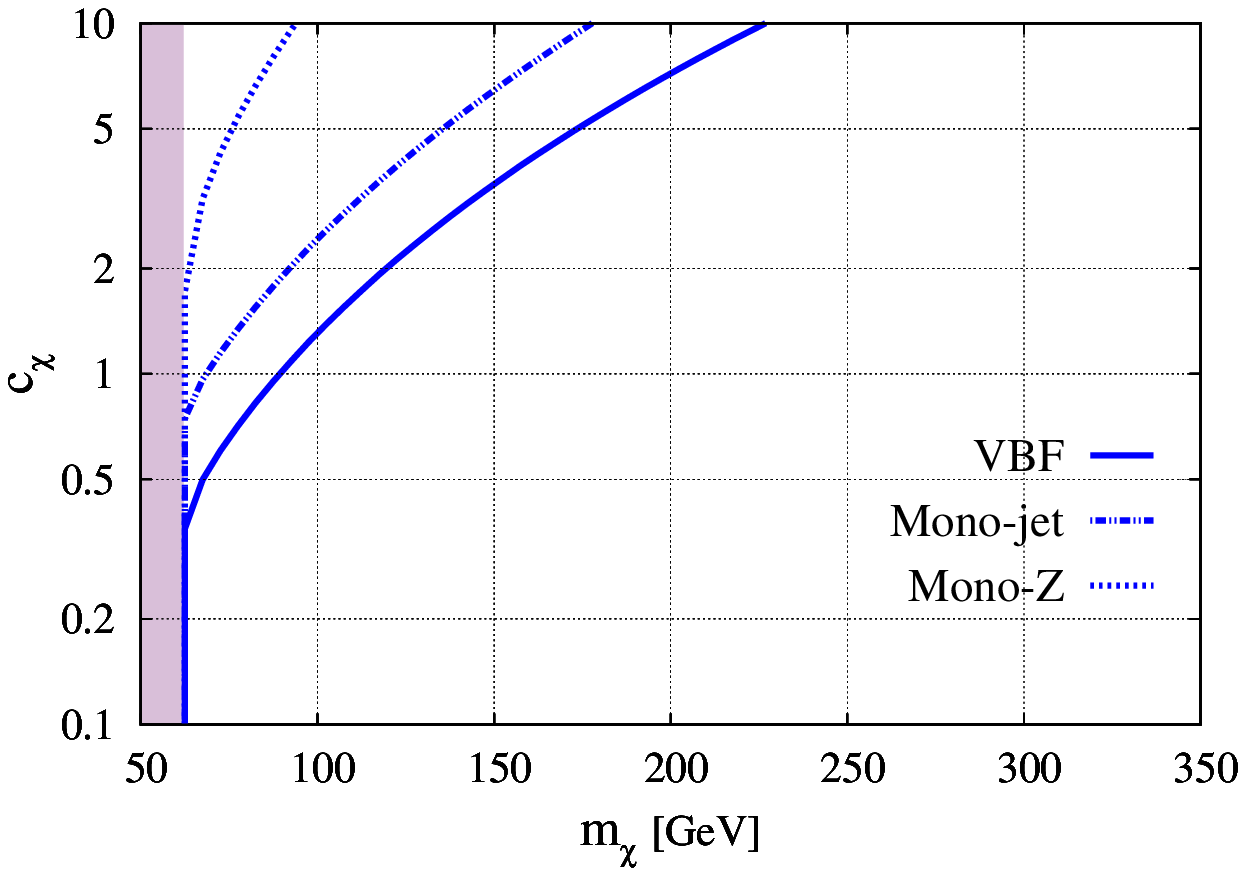}
}
\subfigure[]{
 \includegraphics[scale=0.6]{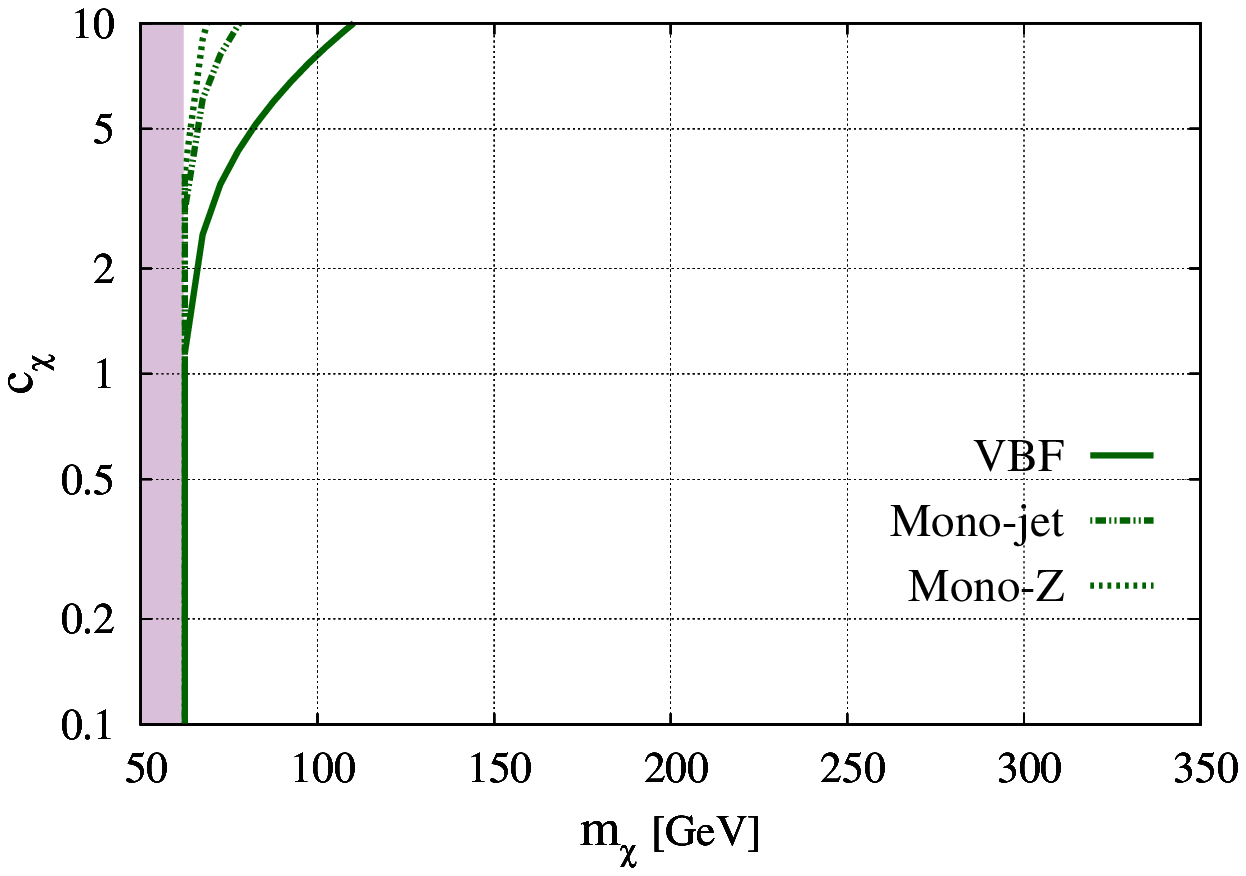}
}
\end{center}                      
 \caption{The 8\,TeV LHC constraints on the tensor (a), vector (b) and
 scalar (c) Higgs-portal couplings,
 $c_\chi$, as a function of the WIMP mass
 $m_\chi$. The solid, dot-dashed and dotted curves correspond to the
 95\% CL upper bounds based on the VBF, mono-jet and mono-$Z$
 data \cite{Chatrchyan:2014tja,CMS:rwa,Aad:2014vka}. The shaded area
 for $m_\chi < m_h/2$ is excluded by the Higgs invisible decay analysis\cite{Aad:2014iia,Chatrchyan:2014tja}.}
\label{fig:Climit}
\end{figure} 
%%%%%%%%%%%%%%%%%%%%%%%%%%%%%%%%%%%%%%%%%%%%%%%%%%%%

In this section we present the constraints on the Higgs-portal
scalar, vector and tensor productions by the VBF, mono-jet and mono-$Z$ 
studies at the 8\,TeV LHC. 

In Figs.~\ref{fig:Climit} we show the constraints on the Higgs-portal
coupling constant, $c_{\chi}$, defined in Eqs.~(\ref{eq:model}) 
as a function of the dark matter mass, $m_\chi$. 
The figures (a), (b) and (c) show the 95\% CL upper bounds for the
tensor, vector and scalar cases, respectively, based on the VBF (solid), 
mono-jet (dot-dashed) and mono-$Z$ (dotted) analyses \cite{Chatrchyan:2014tja,CMS:rwa}.
There is a rapid change in the bounds around $m_\chi = m_h/2\, (\simeq 62.5\,
{\rm GeV})$ due to the on-shell
Higgs pole. The strong constraint on the Higgs invisible decay branching fraction
for $m_\chi < m_h/2$ (the shaded area) is invalid for $m_\chi > m_h/2$.

The VBF result puts about twice stronger bounds than the mono-jet ones,
while the mono-$Z$ bound is weaker than them. 
This is because the VBF signals are clean against backgrounds, while the cross section 
of the mono-$Z$ productions is too small.
It is found that the VBF process constrains the couplings of the tensor, vector and scalar to be smaller than 0.16, 0.43 and 2.0 at 
$m_\chi = 65\, {\rm GeV}$, respectively. 
The bounds becomes weaker for larger $m_\chi$ since the WIMP production 
cross section is proportional to $1/m_\chi^4$. 
However, for the vector and tensor cases, this reduction of the 
cross section is relaxed due to the $m_H^4/m_\chi^4$ enhancement 
with respect to the scalar case in the large WIMP-pair invariant mass 
($m_H$) region, as discussed in Sec.~\ref{sec:Model}. Especially, 
the tensor coupling receives the most stringent constraint in the all 
range of the WIMP mass, which is stronger than the vector case by a
factor of $\sim 3$. If the coupling constant becomes as large as $c_\lambda \sim 
4\pi \sim 10$, the perturbative description of Eq.~\eqref{eq:xsec}
breaks down. Thus, the VBF, mono-jet and mono-$Z$ bounds work only below
$m_\chi \sim 320$, 250 and 130\,GeV for the tensor interaction, 
while the limit is 220, 170 and 90\,GeV and 110, 80 and 70\,GeV for the 
vector and scalar ones, respectively.

Although these collider limits are weaker than that from the relic
abundance and direct detection of the DM \cite{Cata:2014sta,Kanemura:2010sh}, we 
stress that the collider limit is valid even if the hidden sector 
particle of interest is not the dominant component of the DM.

%%%%%%%%%%%%%%%%%%%%%%%%%%%%%%%%%%%%%%%%%%%%%%%%%%%%
\section{Future Prospects}
\label{sec:prospects}
In this section, we discuss future prospects of the sensitivity 
to the Higgs-portal models at the 14\,TeV LHC. 

\subsection{Mono-jet prospects}

%%%%%%%%%%%%%%%%%%%%%%%%%%%%%%%%%%%%%%%%%%%%%%%%%%%%
\begin{figure}[tbp]
\begin{center}
\includegraphics[scale=0.8]{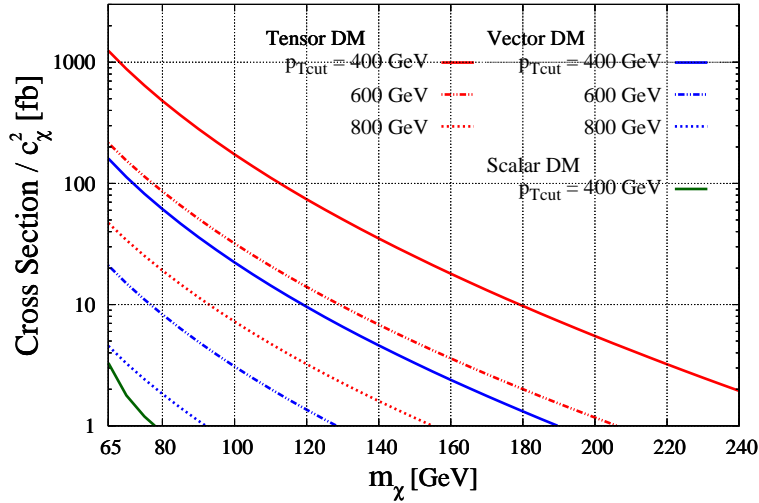}
\end{center}
 \caption{The cross sections of the tensor, vector and scalar WIMP-pair productions associated with
 one jet at the 14\,TeV LHC as functions of the WIMP mass. The vertical axis is rescaled by the Higgs-portal
 coupling squared. The red, blue and green curves
 are for the tensor, vector and scalar cases, respectively, with the $p_{TH}$ cut of 400 (solid), 600
 (dot-dashed) and 800\,GeV (dotted).}
\label{fig:xsec14LHC}
\end{figure}
%%%%%%%%%%%%%%%%%%%%%%%%%%%%%%%%%%%%%%%%%%%%%%%%%%%%

In order to discuss future prospects, we need to evaluate the signal event number and backgrounds. 
First, let us estimate expected backgrounds and upper bounds on the signal excess with the following assumptions:
\begin{itemize}
 \item 
 Backgrounds event number ($N_{\rm BG}$) is scaled by 
 the integrated luminosity,
 $\mathcal{L}$, and the parton luminosity function, $L(\hat{s})$, at a
       typical center-of-mass energy scale of partonic collisions,
       $\sqrt{\hat{s}}$, for
       the background process. Here the parton luminosity function is
       defined as
 \begin{align}
L(\hat{s}) \equiv
  \int^{{\rm ln}(1/\sqrt{\tau})}_{-{\rm ln}(1/\sqrt{\tau})} dy\,
  f_a(\sqrt{\tau} e^y,Q) f_b(\sqrt{\tau}e^{-y},Q), 
 \hspace{1em} \tau = \frac{\hat{s}}{s},
 \end{align}
where $f_i(x,Q)$ is a PDF for an incoming parton flavor $i$ with a
       longitudinal momentum fraction $x$ at a
       factorization scale $Q$.
 Thus, from the experimental result at $\sqrt{s}=8$\,TeV,
 $N_{\rm BG}$ at $\sqrt{s}=14$\,TeV is deduced as
 \begin{align}
   N_{\rm BG}^{14\,{\rm TeV}} = 
  \frac{\mathcal{L}^{14\,{\rm TeV}}}{\mathcal{L}^{8\,{\rm TeV}}}
   \frac{L^{14\,{\rm TeV}}(\hat s)}{L^{8\,{\rm TeV}}(\hat s)}
   N_{\rm BG}^{8\,{\rm TeV}},
 \end{align}
% where $\hat s$ is a typical center-of-mass energy scale of the partonic collision in the background process.
 For the mono-jet process, the background is dominated by $Z + {\rm jets}$ 
 productions. Then, the parton luminosity function is calculated for $q\bar q$
 initial states $(q = u,d,s,c,b)$ with $Q = \sqrt{\hat{s}} = \sqrt{m_Z^2 +\slashed{p}_{T {\rm cut}}^2} + \slashed{p}_{T {\rm cut}}$,
 where $\slashed{p}_{T {\rm cut}}$ is the missing $p_T$ cut. 

 \item 
 The relative systematic uncertainty is the same level as those at the 8\,TeV LHC, 
 \begin{align}
   \frac{\sigma^{14\,{\rm TeV}}_{\rm sys}}{N_{\rm BG}^{14\,{\rm TeV}}} = 
   \frac{\sigma^{8\,{\rm TeV}}_{\rm sys}}{N_{\rm BG}^{8\,{\rm TeV}}},
 \end{align}
 while the statistical uncertainty is scaled as
 \begin{align}
   \frac{\sigma^{14\,{\rm TeV}}_{\rm stat}}{\sqrt{N_{\rm BG}^{14\,{\rm TeV}}}} = 
   \frac{\sigma^{8\,{\rm TeV}}_{\rm stat}}{\sqrt{N_{\rm BG}^{8\,{\rm TeV}}}}.
 \end{align}
 The total uncertainty $\sigma_{\rm tot}$ is estimated by adding these uncertainties in quadrature.

 \item 
 Expected upper bound on the signal excess is estimated as
 \begin{align}
   N_{\rm sig} < 2\, \sigma_{\rm tot}.
 \end{align}
\end{itemize}
Next, in Fig.~\ref{fig:xsec14LHC} we show the cross sections 
of the jet-associated WIMP-pair production process, $pp\rightarrow \chi\chi j$, 
at the 14\,TeV LHC as functions of the WIMP mass. 
Here, the vertical axis is rescaled by the Higgs-portal coupling squared, $c_\chi^2$,
since the cross section is proportional to $c_\chi^2$.
The red, blue and green curves are for the tensor, vector and
scalar cases, respectively.
The cut conditions are supposed to be the same as the 8\,TeV analysis 
(see Eq.~\eqref{eq:cut-monojet}), but the $p_{TH}$ cut is taken as
400 (solid), 600 (dot-dashed) and 800\,GeV (dotted).

The estimated upper bound becomes $\sim 4900$ events for the
$p_{TH}$ cut of 400\,GeV at ${\cal L} = 100 \invfb$, based on the CMS mono-jet
result at the 8\,TeV LHC\cite{CMS:rwa}. 
On the other hand, the signal event number for the tensor production with
$m_\chi=65$\,GeV is estimated to be $\sim 1200$ for $c_\chi=0.1$. 
Thus, if the $p_{TH}$ cut of 400\,GeV is still applicable at the 14\,TeV LHC, 
the Higgs-portal tensor model can be probed for a coupling as small as
$\sim 0.2$ at 
$m_\chi=65$\,GeV and ${\cal L} = 100\,\invfb$. 
Expected signal event number decreases rapidly for heavier WIMPs. 
The signal becomes about 10 events for $m_\chi=180$\,GeV with $c_\chi = 0.1$ 
at ${\cal L} = 100\,\invfb$.  

For $p_{TH} > 550$\,GeV, the expected signal
events and the expected upper bound on the signal excess become $\sim
300$ (for $c_\chi = 0.1$) and $\sim 2000$ events at ${\cal L}
= 100\,\invfb$, respectively. This leads to a weaker limit on the Higgs-portal coupling as
$\sim 0.3$. Thus, harder $p_{TH}$ cut may degrade the
sensitivity to the Higgs-portal couplings for the mono-jet channel at the
14\,TeV LHC. Dedicated studies on the backgrounds and cut conditions are needed 
for a definitive conclusion. 

The sensitivities to the vector and scalar models are weaker than the tensor one. 
The production cross section of the vector is smaller by a factor than
that of the tensor, and the tensor result can be reinterpreted to the vector model straightforwardly. 
On the other hand, the mono-jet channel will be hopeless to probe the Higgs-portal 
scalar model with a coupling of $\sim 0.1$ even at the 14\,TeV LHC because of the too small production cross section.

%%%%%%%%%%%%%%%%%%%%%%%%%%%%%%%%%%%%%%%%%%%%%%%%%%%%
\subsection{Mono-$Z$ prospects}

%%%%%%%%%%%%%%%%%%%%%%%%%%%%%%%%%%%%%%%%%%%%%%%%%%%%
\begin{figure}[tbp]
\begin{center}
\includegraphics[scale=0.8]{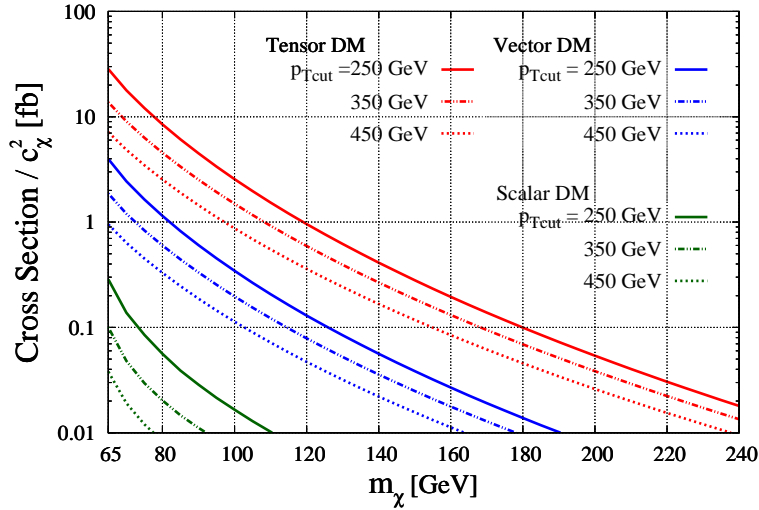}
\end{center}
 \caption{The cross sections of the tensor, vector and scalar WIMP-pair productions associated with
 a Z boson at the 14\,TeV LHC as functions of the WIMP mass. 
 The vertical axis is rescaled by the Higgs-portal
 coupling squared. The red, blue and green curves
 are for the tensor, vector and scalar cases, respectively, with the $p_{TH}$ cut of 250 (solid), 350
 (dot-dashed) and 450\,GeV (dotted).}
\label{fig:xsecMonoZ14LHC}
\end{figure}
%%%%%%%%%%%%%%%%%%%%%%%%%%%%%%%%%%%%%%%%%%%%%%%%%%%%

Future prospects at the 14\,TeV LHC for the mono-$Z$ channel is discussed along the
lines of the previous subsection for the mono-jet channel.
Expected backgrounds and upper bounds on the signal excess are estimated in a similar 
way, but the dominant background is $ZZ$ productions. 
Then, the parton luminosity function is calculated for $q\bar q$ initial
states $(q = u,d,s,c,b)$ with 
$Q=\sqrt{\hat{s}} = 2\sqrt{m_Z^2 + \slashed{p}_{T {\rm cut}}^2}$. 

In Fig.~\ref{fig:xsecMonoZ14LHC}, we show the cross sections of the $Z$-associated WIMP-pair production process, $pp\rightarrow \chi\chi Z$, at the 14\,TeV LHC as functions of the WIMP mass. 
The red, blue and green curves show the cross sections for the tensor, vector and
scalar cases, respectively.
Here, the cut conditions are supposed to be the same as the 8\,TeV analysis, 
Eq.~\eqref{eq:cut-monoZ}, but the $p_{TH}$ cut is set as 250 (solid), 
350 (dot-dashed) or 450\,GeV (dotted). 
The event number of the tensor production with $m_\chi=65$\,GeV is estimated to be 
$\sim 7$ events for $c_\chi=0.1$ and $p_{TH} > 450$\,GeV at ${\cal L}=100\,\invfb$, 
while the expected upper bound on the signal excess for this integrated luminosity 
is $\sim 15$ events, deduced from the ATLAS mono-$Z$ result at the 8\,TeV LHC\cite{Aad:2014vka}. 
Thus, if the $p_{TH}$ cut of 450\,GeV is applicable at the 14\,TeV LHC, 
the Higgs-portal tensor model can be probed for a coupling as small as $\sim 0.15$ at 
$m_\chi=65$\,GeV and ${\cal L} = 100\,\invfb$. It is mentioned that the backgrounds 
are expected to be suppressed more efficiently by adopting harder $p_{TH}$ cut in the 
mono-$Z$ channel at the 14\,TeV LHC. This contrasts sharply with the mono-jet case.

The mono-$Z$ sensitivities to the vector model are weaker than the tensor case. 
On the other hand, the mono-$Z$ channel will be hopeless to probe the scalar model 
 with a coupling of $\sim 0.1$ even in the 14\,TeV LHC as the mono-jet channel. 

%%%%%%%%%%%%%%%%%%%%%%%%%%%%%%%%%%%%%%%%%%%%%%%%%%%%
\subsection{VBF prospects}
Let us turn to the VBF process. 
Future prospects of the Higgs invisible decay branching fraction for
the process have been discussed in 
Ref.~\cite{Ghosh:2012ep}. 
The current upper bound is $\brinv < 0.65$ for $m_h=125$\,GeV, 
and the sensitivity is expected to be $\brinv \sim 0.17$ for $\sqrt{s}=14$\,TeV and $\mathcal{L}=100\,\invfb$. 
Thus, the limit on $\sigma \times \brinv$ will be improved by 
a factor of $3.8$ for $m_h=125$\,GeV.
If this factor is assumed to be independent of the Higgs boson mass and the 
kinematical distributions, the limit on the Higgs-portal WIMP coupling, $\clim$, 
is expected to be improved by a factor of $\sim 2$, since the WIMP production cross section, Eq.~\eqref{eq:xsec}, is proportional to $c_\chi^2$.
Therefore, the expected limit is estimated to be $\clim \sim 0.08$, 0.2 and 1.0 
for the tensor, vector and scalar cases, respectively, at $m_\chi = 65$\,GeV. 
It should be noted that careful background estimations with reliable detector 
simulations are required to derive more realistic conclusions on 14\,TeV 
prospects for the Higgs-portal interactions.\footnote
{
The background simulation in the VBF analysis~\cite{Ghosh:2012ep} seems 
to discard some contributions such as those due to pileup effects.
The signal significance may be degraded especially when the luminosity is large. 
}

%%%%%%%%%%%%%%%%%%%%%%%%%%%%%%%%%%%%%%%%%%%%%%%%%%%%
\section{Conclusion}
\label{sec:conclusion}
We discussed the LHC constraints on the several Higgs-portal WIMP
models at the LHC. WIMPs are assumed to be scalar, vector and anti-symmetric tensor fields. 
In particular, we considered their masses heavier than a half of 
the Higgs boson mass, where LHC analyses have been rarely performed.
We found that the current LHC data excludes the models with the Higgs-portal 
couplings larger than 
0.16 (tensor), 0.43 (vector) and 2.0 (scalar) for $m_\chi=65$\,GeV, 
while these constraints become weaker for heavier WIMPs. 
The tensor and vector Higgs-portal couplings receive
stronger limits than the scalar one especially when WIMPs are 
heavy. The tensor WIMP coupling is constrained most strongly for
any WIMP masses. On the other hand, for $m_\chi > 320$ (tensor), $220$
(vector) and $110$\,GeV (scalar), our analyses cannot be applied since the perturbative calculations break down.

We also discussed the prospects for 14\,TeV LHC sensitivities to the
Higgs-portal interactions. 
For the mono-jet and mono-$Z$ processes,  
we evaluated the production cross sections of the Higgs-portal WIMPs with 
several cut conditions. 
It was shown that for mono-jet (mono-$Z$) process more than 1000 (5) events could be produced for the tensor case
with $c_\chi = 0.1$ at ${\cal L} = 100\,\invfb$ for $p_{TH} > 400\,
(450)$\,GeV.
Although it is difficult to simulate the backgrounds 
and to estimate the signal significance reliably, we roughly estimated
the 14\,TeV LHC reach for the Higgs-portal couplings by deducing
the upper limits on the signal excesses based on the 8\,TeV LHC
results. It is found that both mono-jet and mono-$Z$ channels will
improve their limits, though this improvement depends on the
missing $p_T$ cut condition. The mono-jet channel may prefer the current level
cut, while the harder cut will be helpful for the mono-$Z$ channel.
It would be still challenging to constrain the scalar WIMP effectively in the 
$m_\chi > m_H/2$ region even for the 14\,TeV LHC. 
Besides, the Higgs invisible decay channel is expected to be 
studied accurately by the VBF signal. 
Then, we may expect a factor of two improvement of the coupling limits at $\mathcal{L}=100\,\invfb$. 
Dedicated studies on realistic estimations of the 
backgrounds and detector effects are needed for further discussions.

As discussed in this letter, the LHC can probe WIMPs through the Higgs portal not only when they are lighter than a half of the Higgs boson 
mass but also if they are heavier. 
Therefore, the LHC will be useful in searching for the Higgs-portal 
models for wider cases than had been expected.

%%%%%%%%%%%%%%%%%%%%%%%%%%%%%%%%%%%%%%%%%%%%%%%%%%%%
\subsection*{Acknowledgements}
This work was supported by JSPS KAKENHI Grant No.~23740172 (M.E.) and
No.~60322997  (Y.T.).

%%%%%%%%%%%%%%%%%%%%%%%%%%%%%%%%%%%%%%%%%%%%%%%%%%%%
\providecommand{\href}[2]{#2}
\begingroup\raggedright

\endgroup
\end{document}